\documentclass[prd,reprint,showpacs,showkeys]{revtex4-1}
\usepackage{amsfonts}
\usepackage{mdframed}
\usepackage{amssymb}
\usepackage{amsmath}
\usepackage{graphicx}
\usepackage[font={footnotesize,it}]{caption}

\begin{document}

\preprint{}
\title{Generation of spherically symmetric metrics in $f\left( R\right) $
gravity }
\author{Z. Amirabi}
\email{zahra.amirabi@emu.edu.tr}
\author{M. Halilsoy}
\email{mustafa.halilsoy@emu.edu.tr}
\author{S. Habib Mazharimousavi}
\email{habib.mazhari@emu.edu.tr}
\affiliation{Department of Physics, Eastern Mediterranean University, Gazima\u{g}usa,
Turkey.}

\begin{abstract}
In $D-$dimensional spherically symmetric $f\left( R\right) $ gravity there
are three unknown functions to be determined from the fourth order
differential equations. It is shown that the system remarkably integrates to
relate two functions through the third one to provide reduction to second
order equations accompanied with a large class of potential solutions. The
third function which acts as the generator of the process is $F\left(
R\right) =\frac{df\left( R\right) }{dR}.$ We recall that our generating
function has been employed as a scalar field with an accompanying
self-interacting potential previously which is entirely different from our
approach. Reduction of $f\left( R\right) $ theory into system of equations
seems to be efficient enough to generate a solution corresponding to each
generating function. As particular examples, besides known ones, we obtain
new black hole solutions in any dimension $D$. We further extend our
analysis to cover non-zero energy-momentum tensors. Global monopole and
Maxwell sources are given as examples.
\end{abstract}

\pacs{}
\maketitle

\section{Introduction}

$f\left( R\right) $ gravity is one of the modified theories of Einstein's
general relativity that attracted much attention in recent times \cite%
{1,2,3,4}. In \cite{5} $f\left( R\right) =R+\alpha R^{2}$ with $\alpha >0$
has been introduced as the model of inflated universe while $f\left(
R\right) =R-\alpha /R^{n}$ ($\alpha >0$, $n>0$) was considered as a
candidate for the dark energy model \cite{6,7,8,9,10,11}. This model,
however, is not a viable model for dark energy and instead $f\left( R\right)
=R-\alpha R^{n}$ with $\alpha >0$ and $0<n<1$ emerged as alternative which
has been proposed in \cite{12,13}. Later on more viable models were studied
in \cite{14,15,16,17,18}. A detailed review of these models \ are given in 
\cite{19} (Other review papers are given in \cite{20,21,22,23}). Some recent
works on solutions in $f\left( R\right) $ gravity are given in \cite%
{24,25,26,27,28,29,30,31,32,33,34,35,36,37,38,39,40,41,42,43,44,45,46}. For $%
f\left( R\right) =R$, in $D=4$ dimensional spacetime it coincides with the
standard general relativity but otherwise it stands with an action that is
arbitrarily dependent on the Ricci scalar. Finding exact solutions in this
theory with fourth order derivatives of the metric tensor is both important
and challenging. Apart from exact analytic solutions there are $f\left(
R\right) $ models that can only be expressed implicitly in non-polynomial
expressions. Each particular model has advantages / disadvantages as far as
experimental tests are concerned \cite{47,48,49,50,51,52,53}. There are even
models that lack the Einstein's $R-$gravity limit. Among other expectations
the UV / IR behaviors at near / far distances, quantum renormalizability
with power counting of the counter terms are prominent. At any cost
preferring to abide by the classical regime we confine ourselves first to
sourceless (vacuum) $f\left( R\right) $ models that admits exact integrals.
In the last section we extend our discussion to cover external sources such
as global monopole \cite{54} and electromagnetic field. Let us add that the
equivalence of $f\left( R\right) $ gravity to Brans-Dicke (BD) theory ($%
\omega =0$) with a potential has also been highlighted extensively in the
past as a transition between Jordan and Einstein frames. In this approach
the exact solutions can be generated by adopting scalar field ansatzes which
in general brings into the Lagrangian intricate potentials. Our method will
be confined entirely to the Jordan frame without reference to the BD field
or any scalar potential.

The vacuum of $f\left( R\right) $ gravity is known to carry its own
curvature sources. By vacuum in this theory it is meant the absence of an
external energy-momentum tensor $T_{\mu \nu }$, of any physical source \cite%
{55}. Carames and de Mello in latter work have considered the spherically
symmetric vacuum solutions of $f\left( R\right) $ gravity in higher
dimensions. We shall rederive most of their results anew together with some
additional extensions which we are interested in to develop further.
Addition of external $T_{\mu \nu }\neq 0,$ no doubt makes the problem
technically more complicated, but following the lesson learned from the
vacuum / empty solutions of $f\left( R\right) =R$ theory of gravity we will
attempt to derive the most general equations and in some specific cases the
solutions as well. The $D$ dimensional spherically symmetric line element
that we shall consider will be 
\begin{equation}
ds^{2}=-A\left( r\right) dt^{2}+\frac{1}{B\left( r\right) }%
dr^{2}+r^{2}d\Omega _{D-2}^{2}
\end{equation}%
in which $A\left( r\right) $ and $B\left( r\right) $ are metric functions to
be determined while $d\Omega _{D-2}^{2}$ represents the $\left( D-2\right) -$%
dimensional unit spherical line element. In particular integrals we have the
restrictive case $A\left( r\right) =B\left( r\right) $ included, but more
general cases with $A\left( r\right) \neq B\left( r\right) $ must be
interesting as well. Beside the metric functions $A\left( r\right) $ and $%
B\left( r\right) $ we shall employ a third function denoted by $F\left(
R\right) =\frac{df\left( R\right) }{dR},$ which characterizes the type of
the $f\left( R\right) $ gravity. Let us add that not in all cases of $%
f\left( R\right) $ models the explicit form of $f\left( R\right) $ can be
expressed analytically in terms of the variable $R,$ the Ricci scalar.
Instead, it involves a transcendental part that can't be inverted in the
form of $r=r\left( R\right) ,$ these are hybrid forms. We add that even
these hybrid forms don't prevent us from calculating $\frac{df}{dR}>0$ and $%
\frac{d^{2}f}{dR^{2}}>0$ which are crucial terms to determine the absence of
ghosts and thermodynamic stability, respectively.

In brief what has been achieved in this paper is to show that the metric
functions $A\left( r\right) $ and $B\left( r\right) $ are related through an
integral expression for the function $F\left( R\right) $ (or $F\left(
r\right) $). This amounts to the fact that once $F\left( r\right) $ is given
it acts as a generator to generate a new set of ($A\left( r\right) $, $%
B\left( r\right) $) pair. The function $A\left( r\right) $ is expressed in
terms of $B\left( r\right) $ and $F\left( r\right) $ and the remaining
equation is reduced into a master equation satisfied by $B\left( r\right) $.
Once we give an ansatz for $F\left( r\right) $ our master equation can be
integrated in principle to obtain $B\left( r\right) .$ In this manner we can
obtain an infinite class of metrics in $f\left( R\right) $ gravity generated
from an infinite set of $F\left( r\right) .$ No doubt the dimensionality of
spacetime $D$($=d+1$) also plays role in the derivation. In particular, we
present examples of new black hole solutions in $D\geq 3,$ by the method
described above. We wish to add also that in the reduction process the
system of differential equations in $f\left( R\right) $ gravity reduce
naturally from the fourth order to the second order.

The paper is organized as follows. In Section II we rederive the $f\left(
R\right) $ field equations in $D-$dimensions which is comparable in some
sense with \cite{55}. A number of examples to justify the effectiveness of
our method are given. Generalization to $T_{\mu \nu }\neq 0$ is analyzed in
Section III. We end our discussion with Conclusion in Section IV.

\section{The field equations in $D-$dimensions}

The $D-$dimensional vacuum $f\left( R\right) -$gravity is represented by the
action%
\begin{equation}
I=\frac{1}{16\pi G}\int d^{D}x\sqrt{-g}f\left( R\right)
\end{equation}%
in which $f\left( R\right) $ is a function of Ricci scalar $R$ and $D\geq 3$%
. Variation of the action $I$ with respect to $g_{\mu \nu }$ provides the
field equations (in metric formalism)%
\begin{equation}
FR_{\mu }^{\nu }-\frac{1}{2}f\delta _{\mu }^{\nu }-\nabla ^{\nu }\nabla
_{\mu }F+\delta _{\mu }^{\nu }\square F=0
\end{equation}%
in which $F=\frac{df}{dR}$ and $\square $ is the covariant Laplacian. The
general spherically symmetric line element is given by (1) and the field
equations (3) are explicitly given by%
\begin{equation}
FR_{t}^{t}-\frac{f}{2}+B\left( F^{\prime \prime }+\frac{B^{\prime }F^{\prime
}}{2B}\right) +\frac{D_{2}BF^{\prime }}{r}=0,
\end{equation}%
\begin{equation}
FR_{r}^{r}-\frac{f}{2}+\frac{BA^{\prime }F^{\prime }}{2A}+\frac{%
D_{2}BF^{\prime }}{r}=0
\end{equation}%
and%
\begin{equation}
FR_{\theta }^{\theta }-\frac{f}{2}+\frac{BF^{\prime }}{2}\left( \frac{%
A^{\prime }}{A}+\frac{B^{\prime }}{B}+\frac{2D_{3}}{r}\right) +BF^{\prime
\prime }=0
\end{equation}%
in which%
\begin{equation}
R_{t}^{t}=-\frac{1}{4A}\left( B^{\prime }A^{\prime }+2BA^{\prime \prime }-%
\frac{BA^{\prime 2}}{A}+\frac{2D_{2}BA^{\prime }}{r}\right) ,
\end{equation}%
\begin{equation}
R_{r}^{r}=-\frac{1}{4A}\left( B^{\prime }A^{\prime }+2BA^{\prime \prime }-%
\frac{BA^{\prime 2}}{A}+\frac{2D_{2}B^{\prime }A}{r}\right)
\end{equation}%
and%
\begin{equation}
R_{\theta _{i}}^{\theta _{i}}=-\frac{1}{2r^{2}A}\left( rBA^{\prime
}+rAB^{\prime }+2D_{3}A\left( B-1\right) \right) ,
\end{equation}%
for $1\leq i\leq D_{2}.$ Herein and in the rest of the paper, $D_{k}=D-k$, a
prime stands for the derivative with respect to $r$ and $F=\frac{df}{dR}.$ A
particular combination of these three equations leads to two equations which
are independent of $f$. The first equation (4) may be written as 
\begin{equation}
2ABrF^{\prime \prime }+H\left( AB^{\prime }-BA^{\prime }\right) =0
\end{equation}%
which integrates to%
\begin{equation}
A=BH^{2}\exp \left( -2D_{1}\int \frac{F^{\prime }}{H}dr\right)
\end{equation}%
where 
\begin{equation}
H=rF^{\prime }+D_{2}F\neq 0.
\end{equation}%
The second equation (5) upon considering (11) becomes independent of $A$
too. The closed form of the second equation reduces to a linear equation for 
$B\left( r\right) ,$ which is given by%
\begin{equation}
B^{\prime \prime }+PB^{\prime }+2QB+\frac{2D_{3}}{r^{2}}=0,
\end{equation}%
where%
\begin{equation}
P=\frac{3r}{H}F^{\prime \prime }+\frac{H}{rF}-\frac{2}{r}
\end{equation}%
and%
\begin{equation}
Q=\frac{rF^{\prime \prime \prime }}{H}+\frac{\left( H^{2}-D_{1}rFF^{\prime
}\right) F^{\prime \prime }}{FH^{2}}+\frac{F-H}{r^{2}F}.
\end{equation}%
Finally, the explicit form of $f$ in terms of $r$ is given by%
\begin{multline}
f=\frac{2B\left( 2rF^{\prime }+FD_{3}\right) F^{\prime \prime }}{H} \\
+\frac{2rF^{\prime }\left( rB^{\prime }+D_{3}B\right) -2\left( rB^{\prime
}+D_{3}\left( B-1\right) \right) F}{r^{2}}.
\end{multline}%
To complete our analysis we give the explicit form of the Ricci scalar in
terms of $r$ which reads%
\begin{multline}
R=-B^{\prime \prime }-\frac{D_{2}\left( 2rB^{\prime }+D_{3}\left( B-1\right)
\right) }{r^{2}} \\
-\frac{2r^{3}BF^{\prime \prime \prime }}{Hr^{2}}+\frac{r^{2}F^{\prime \prime
}\left( 3r^{2}B^{\prime }F^{\prime }+FD_{2}\left( 3rB^{\prime
}+2BD_{1}\right) \right) }{H^{2}r^{2}}.
\end{multline}

In summary, the only equation to be solved is Eq. (13) which is second order
and linear for $B$. Therefore the procedure is reduced to set a $F\left(
r\right) $ - which eventually represents the form of $f\left( R\right) $ -
and solve the only equation (13) to find $B\left( r\right) $ and
consequently $A\left( r\right) .$

\subsection{Applications of the method}

Before we give certain applications for our formalism we would like to
compare our approach with the work of Carames and Bezerra de Mello \cite{55}%
. The main difference can be seen from the fact that in \cite{55} there are
two generating functions (so to say) which are $F$ and $Y$. In other words
Eqs. (16) and (17) of \cite{55} are coupled and one must consider them
together to find a solution to the field equations. In our formalism we have
only one generating function which is $F$ and Eq. (13) is the only equation
to be solved. In the following three cases we shall show that for simple
cases ($F=1$ and $F=1+\alpha r$) our results overlap with \cite{55} but for
more complicated case ($F=\alpha r^{a}$) our solution is the general one
while the solution given in \cite{55} is a restricted one (look at Eq. (46)
in \cite{55} and (35) and (36) in this work).

\subsubsection{$F\left( r\right) =1$}

We start with the simplest case, with $F=1$ or equivalently $H=D_{2}.$
Definition of $F=\frac{df}{dR}$ implies 
\begin{equation}
f=R-2\Lambda
\end{equation}%
in which $-2\Lambda $ is an integration constant to be interpreted as the
cosmological constant. The main equations (11) and (13) admit%
\begin{equation}
B\left( r\right) =A\left( r\right) =\left\{ 
\begin{array}{cc}
1+\frac{C_{1}}{r^{D_{3}}}+C_{2}r^{2}, & D>3 \\ 
C_{1}+C_{2}r^{2}, & D=3%
\end{array}%
\right.
\end{equation}%
for integration constants $C_{1}$ and $C_{2}$ and consequently%
\begin{eqnarray}
f\left( r\right) &=&-2C_{2}D_{1}, \\
R &=&-C_{2}DD_{1}  \notag
\end{eqnarray}%
with%
\begin{equation}
\Lambda =-\frac{1}{2}D_{2}D_{1}C_{2}.
\end{equation}%
Finally the solution becomes%
\begin{equation}
B\left( r\right) =A\left( r\right) =\left\{ 
\begin{array}{cc}
1-\frac{2M}{D_{3}r^{D_{3}}}-\frac{2\Lambda }{D_{2}D_{1}}r^{2}, & D>3 \\ 
-M-\Lambda r^{2}, & D=3%
\end{array}%
\right.
\end{equation}%
in which we set 
\begin{equation*}
C_{1}=\left\{ 
\begin{array}{cc}
-\frac{2M}{D_{3}} & D>3 \\ 
-M & D=3%
\end{array}%
\right.
\end{equation*}%
where $M$ is the ADM mass. The solution for $D>3$ is Schwarzschild de / anti
de-Sitter black hole solution and for $D=3$ it is the BTZ black hole. Let's
add that $F\left( r\right) =\xi =$constant, does not change the nature of
the solution for the metric with constant scalar curvature. We note that our
results in this section expectedly is the same as section 3.1 in \cite{55}.

\subsubsection{$F\left( r\right) =1+\protect\alpha r$, with $\protect\alpha %
= $constant}

As of Sec. 3.2 of Ref. \cite{55} we consider $F^{\prime \prime }\left(
r\right) =0$ in our general formalism but instead of going through a general 
$D-$dimensional solution, we investigate the cases in closed form for $%
f\left( R\right) .$ From Eq. (10) with $F^{\prime \prime }=0$ one obtains $%
A=B$ (up to a constant which one can set it unity via a redefinition of
time). For an arbitrary $D$ the solution for $B$ may not be possible in a
closed form but for specific dimensions we may find.

\paragraph{$D=3$}

In three dimensional spacetime the solution for $B$ becomes%
\begin{equation}
B=A=C_{2}r^{2}+C_{1}\left( \alpha r-\frac{1}{2}\right) -C_{1}r^{2}\alpha
^{2}\ln \left( 1+\frac{1}{\alpha r}\right)
\end{equation}%
and 
\begin{equation}
f=-4C_{2}+4C_{1}\alpha ^{2}\ln \left( 1+\frac{1}{\alpha r}\right) -\frac{%
2C_{1}\alpha \left( 1+2\alpha r\right) }{r\left( 1+\alpha r\right) }
\end{equation}%
where $C_{1}$ and $C_{2}$ are integration constants with the curvature
scalar 
\begin{equation}
R=-6C_{2}+6C_{1}\alpha ^{2}\ln \left( 1+\frac{1}{\alpha r}\right) -\frac{%
C_{1}\alpha \left( 2+9\alpha r+6\alpha ^{2}r^{2}\right) }{r\left( 1+\alpha
r\right) ^{2}}.
\end{equation}%
The hybrid relation between $f\left( r\left( R\right) ,R\right) $ and $R$ is
given by%
\begin{equation}
f=\frac{2}{3}R-\frac{2C_{1}\alpha }{3r\left( 1+\alpha r\right) ^{3}}.
\end{equation}%
Note also that from the foregoing expressions we can identify the
cosmological constant by $C_{2}=-\Lambda .$\ Furthermore, setting $\alpha =0$%
, one recovers the previous example with $C_{1}$ scaled which suggests that
it is related to the mass of the central object.

\paragraph{$D=4$}

In four dimensions the solution is given by%
\begin{multline}
B=A=C_{2}r^{2}+\frac{1}{2}+\frac{1}{3\alpha r} \\
+\frac{C_{1}}{r}\left( 3\alpha r-2-6\alpha ^{2}r^{2}+6\alpha ^{3}\ln \left(
1+\frac{1}{\alpha r}\right) \right)
\end{multline}%
with%
\begin{multline}
f=-6C_{2}-36C_{1}\alpha ^{3}\ln \left( 1+\frac{1}{\alpha r}\right) \\
+\frac{6\alpha C_{1}\left( -1+6\alpha ^{2}r^{2}+3\alpha r\right) }{%
r^{2}\left( 1+\alpha r\right) }+\frac{2\alpha r+1}{r^{2}}
\end{multline}%
and%
\begin{multline}
R=-12C_{2}-72C_{1}\alpha ^{3}\ln \left( 1+\frac{1}{\alpha r}\right) \\
+\frac{6\alpha C_{1}\left( -1+6\alpha ^{2}r^{2}+6\alpha r\right) \left(
1+2\alpha r\right) }{r^{2}\left( 1+\alpha r\right) ^{2}}+\frac{1}{r^{2}}.
\end{multline}%
We see clearly the role of $C_{2}=-\frac{\Lambda }{3}$ and we wish to
proceed with $C_{1}=0$ which implies%
\begin{equation}
B=A=-\frac{\Lambda }{3}r^{2}+\frac{1}{2}+\frac{1}{3\alpha r},\text{ \ \ }%
\alpha \neq 0
\end{equation}%
so that%
\begin{equation}
f=R+2\alpha \sqrt{R-4\Lambda }-2\Lambda .
\end{equation}

This is a black hole solution with a singularity at $r=0$ such that 
\begin{equation}
R=\frac{1}{r^{2}}+4\Lambda .
\end{equation}%
If we set $\Lambda =0,$ with $\alpha <0$ we may introduce a horizon for the
solution located at%
\begin{equation}
r_{h}=\frac{2}{3\left\vert \alpha \right\vert }
\end{equation}%
while for $\alpha >0$ the solution possesses a naked singularity at $r=0.$
To complete our investigation let's determine the absence of ghosts and
thermodynamic stability of the explicit $f(R)$ found in (31). We see that $%
\frac{df}{dR}=1+\frac{\alpha }{\sqrt{R-2\Lambda }}$ and $\frac{d^{2}f}{dR^{2}%
}=-\frac{\alpha }{2\left( R-2\Lambda \right) ^{3/2}}.$ Clearly both
conditions can not be satisfied simultaneously.

\subsubsection{$F\left( r\right) =\protect\alpha r^{a}$}

Our next example is a power-law form for $F\left( r\right) ,$ i.e., 
\begin{equation}
F=\alpha r^{a}
\end{equation}%
with constants $\alpha $ and $a$ which upon (11) yields%
\begin{equation}
A=r^{\frac{2a\left( a-1\right) }{a+D_{2}}}B.
\end{equation}%
Substituting into (13) one finds%
\begin{multline}
B=C_{1}r^{-\frac{2a^{2}-6a+6+\left( 2a-5\right) D+D^{2}}{a+D_{2}}}+C_{2}r^{%
\frac{2\left( D_{2}+2a-a^{2}\right) }{a+D_{2}}} \\
+\frac{D_{3}\left( a+D_{2}\right) ^{2}}{\left( 2a^{2}-6a+6+\left(
2a-5\right) D+D^{2}\right) \left( D_{2}+2a-a^{2}\right) }
\end{multline}%
in which $C_{1}$ and $C_{2}$ are the integration constants. Following $A$
and $B$ we also find%
\begin{equation}
f=\frac{2\alpha C_{2}D_{1}\left( a-1\right) \left( D_{2}+2a\right) r^{\frac{%
a\left( D-a\right) }{a+D_{2}}}}{a+D_{2}}+\frac{2a\alpha D_{1}D_{3}r^{a-2}}{%
D_{2}+2a-a^{2}}
\end{equation}%
and%
\begin{equation}
R=\frac{C_{2}D_{1}\left( D-a\right) \left( D_{2}+2a\right) }{\left(
a+D_{2}\right) r^{\frac{2a\left( a-1\right) }{a+D_{2}}}}-\frac{%
aD_{1}D_{3}\left( a-2\right) }{\left( D_{2}+2a-a^{2}\right) r^{2}}.
\end{equation}%
We also note that although $\alpha $ and $a$ are two arbitrary constants $a$
must satisfy $a\neq -D_{2}$, $1\pm \sqrt{D_{1}}.$ This is remarkable to
observe that in Sec. 3.3.2 of \cite{55} the same ansatz for $F$ has been
considered but the solutions to the field equations (see Eq. (46)-(49) of 
\cite{55}) are not the same as what we found here is more general. As a
matter of fact our solutions (35)-(38) with $C_{2}=0$ reduce to their
solutions. This shows that reducing the field equations into a master
equation with a single generating function makes advances in finding exact
solutions in $f\left( R\right) $ gravity.

In $D=3$ dimensions the solution becomes rather specific since the last term
vanishes for all values of $a.$ The functions then read as 
\begin{equation}
A=r^{\frac{2a\left( a-1\right) }{a+1}}B,
\end{equation}%
\begin{equation}
B=C_{1}r^{\frac{-2a^{2}}{1+a}}+C_{2}r^{\frac{2\left( 1+2a-a^{2}\right) }{a+1}%
},
\end{equation}%
\begin{equation}
f=\frac{4\alpha C_{2}\left( 2a^{2}-a-1\right) }{a+1}r^{\frac{a\left(
3-a\right) }{a+1}}
\end{equation}%
and%
\begin{equation}
R=\frac{2\left( 2a+1\right) \left( a-3\right) C_{2}}{a+1}r^{\frac{2a\left(
1-a\right) }{a+1}}.
\end{equation}%
This is nothing but the solution found by Zhang, Liu and Li in \cite{56}
with their parameters 
\begin{equation}
p=-\frac{2a\left( 1-a\right) }{a+1}
\end{equation}%
and 
\begin{equation}
kL=-\frac{2\left( 2a+1\right) \left( a-3\right) C_{1}}{a+1}.
\end{equation}%
Note that 
\begin{equation}
f\sim R^{\frac{\left( 3-a\right) }{2\left( 1-a\right) }}
\end{equation}%
which yields $f\sim R^{2}$ for the specific choice $a=\frac{1}{3}.$

For $D\geq 4$ one may set $C_{2}=0$ and therefore%
\begin{multline}
B=C_{1}r^{-\frac{2a^{2}-6a+6+\left( 2a-5\right) D+D^{2}}{a+D_{2}}} \\
+\frac{D_{3}\left( a+D_{2}\right) ^{2}}{\left( 2a^{2}-6a+6+\left(
2a-5\right) D+D^{2}\right) \left( D_{2}+2a-a^{2}\right) }
\end{multline}%
with an analytic relation for $f\left( R\right) $ given by%
\begin{equation}
f=\bar{\alpha}R^{1-\frac{a}{2}}
\end{equation}%
in which $\bar{\alpha}$ is a constant which can be set to unity (by a fine
choice of $\alpha $). In the case of $a=0$ the theory gives $R-$gravity. The
solution is a black hole with singularity at $r=0.$ Let us also add that
with $C_{1}=0$ and $a=1$ the solution reduces to 
\begin{equation}
ds^{2}=-\xi dt^{2}+\frac{1}{\xi }dr^{2}+r^{2}d\Omega _{D_{2}}^{2}
\end{equation}%
in which%
\begin{equation}
\xi =\frac{D_{3}}{D_{2}}
\end{equation}%
with%
\begin{equation}
R=\frac{D_{3}}{r^{2}}
\end{equation}%
and%
\begin{equation}
f=\sqrt{R}.
\end{equation}%
This represents a global monopole-type solution with a deficit angle.
Another interesting setting is for $a=-2$ and $C_{2}=0$ which upon a proper
choice of $\alpha $ one finds 
\begin{equation}
f\left( R\right) =R^{2}
\end{equation}%
and the solution becomes 
\begin{equation}
B\left( r\right) =C_{1}r^{-\frac{DD_{9}+26}{D_{4}}}-\frac{D_{3}D_{4}^{2}}{%
\left( DD_{9}+26\right) D_{10}}
\end{equation}%
while%
\begin{equation}
A\left( r\right) =r^{\frac{12}{D_{4}}}B\left( r\right) .
\end{equation}%
Clearly $D=10$ and $D=4$ are excluded. $D=4$ is not allowed directly from
(10) where $H=0$ with $F=\frac{\alpha }{r^{2}}$ demanding $AB=0$ which is
not acceptable. For $D=10$ the particular solution can be obtained as%
\begin{equation}
B=C_{2}-\frac{C_{1}}{6r^{6}}-\frac{7}{3}\ln r,
\end{equation}%
and%
\begin{equation}
f=-\frac{4\alpha \left( 9C_{2}-7-21\ln r\right) }{r^{4}}
\end{equation}%
with the Ricci scalar%
\begin{equation}
R=\frac{98+168\ln r-72C_{2}}{r^{2}}.
\end{equation}%
The metric function $A\left( r\right) $ follows accordingly from (11) which
is given in (54).

We comment that $f=\bar{\alpha}R^{1-\frac{a}{2}}$ does not satisfy $\frac{df%
}{dR}>0$ and $\frac{d^{2}f}{dR^{2}}>0$ simultaneously unless we set $a$ to
be negative (Note that $\bar{\alpha}=1$ is needed to have the Einstein
gravity recovered.). For instance with $\bar{\alpha}=1$ and $a=-2$ which
implies $f=R^{2},$ and both conditions are satisfied.

\subsubsection{A new black hole solution in $D=3$}

In \cite{56} where $f\left( R\right) =R^{d+1}$ in three dimensional
spacetime with $d=const.$ has been studied, the solution does not cover the
case $d=-\frac{1}{2}$ which makes $f\left( R\right) =\sqrt{R}.$ This can be
seen from the Eq. (12) of \cite{56} and Eq. (45) (note that $\frac{3-a}{%
2\left( 1-a\right) }=\frac{1}{2}$ has no answer) and in that paper both do
not cover the case $f\left( R\right) =\sqrt{R}$. However, in what follows we
wish to show that the solution $f\left( R\right) =\sqrt{R}$ can easily be
obtained in three dimensions. To do so let's consider 
\begin{equation}
F\left( r\right) =\beta r\exp \left( \alpha r\right)
\end{equation}%
with $\alpha $ and $\beta $ two real constants. This yields%
\begin{equation}
A=\frac{C_{1}}{r}+C_{2}r^{2},
\end{equation}%
\begin{equation}
B=\left( \frac{C_{1}}{r}+C_{2}r^{2}\right) \exp \left( -2\alpha r\right) ,
\end{equation}%
\begin{equation}
f=2\alpha \beta \left( \frac{C_{1}}{r}+4C_{2}r^{2}\right) \exp \left(
-\alpha r\right)
\end{equation}%
and%
\begin{equation}
R=4\alpha \left( \frac{C_{1}}{4r^{2}}+C_{2}r-\frac{3C_{2}}{2\alpha }\right)
\exp \left( -2\alpha r\right) .
\end{equation}%
Note that, $C_{1}$ and $C_{2}$ are two integration constants such that $%
C_{2} $ effectively plays the role of a cosmological constant. The solution
is a black hole with a horizon located at $r_{h}=\left( \frac{-C_{1}}{C_{2}}%
\right) ^{1/3}$ with the condition that $\frac{-C_{1}}{C_{2}}>0.$ Also from $%
R$ we see that the solution is singular if and only if $C_{1}\neq 0.$ In the
sequel we are interested in $C_{2}=0$ which makes the solution to be rather
simple but singular. Accordingly the form of $f$ and $R$ are given by%
\begin{equation}
f=2\alpha \beta C_{1}\frac{\exp \left( -\alpha r\right) }{r},
\end{equation}%
\begin{equation}
R=\alpha C_{1}\left( \frac{\exp \left( -\alpha r\right) }{r}\right) ^{2},
\end{equation}%
which upon tuning the free parameter $\beta $ by $4\alpha \beta ^{2}C_{1}=1$
the form of $f$ becomes%
\begin{equation}
f=\sqrt{R}.
\end{equation}%
Here $\alpha $ and $C_{1}$ are positive constants and the line element
finally reads%
\begin{equation}
ds^{2}=-\frac{r_{0}}{r}dt^{2}+\frac{r}{r_{0}\exp \left( -2\alpha r\right) }%
dr^{2}+r^{2}d\theta ^{2}
\end{equation}%
in which $r_{0}=\frac{1}{4\alpha \beta ^{2}}$ is also a positive constant.

\subsubsection{A general class of solutions in $2+1-$dimensions}

In three dimensional spacetime in addition to what we found by now, we wish
to show that there exists an important class of solutions yet to be
discovered. This specific class, however is a characteristic feature of only
three dimensions. To see this solution let's set $D=3$ in (13) which yields%
\begin{multline}
r^{2}F\left( rA^{\prime }-2A\right) F^{\prime \prime } \\
-r\left[ \left( rA^{\prime \prime }-A^{\prime }\right) F+\left( rA^{\prime
}-2A\right) F^{\prime }\right] \left( rF^{\prime }+F\right) =0
\end{multline}%
where we substituted $B\left( r\right) $ by $A\left( r\right) $ using (11)
i.e.,%
\begin{equation}
B=\frac{Ae^{4\int \frac{F^{\prime }}{rF^{\prime }+F}dr}}{\left( rF^{\prime
}+F\right) ^{2}}.
\end{equation}%
Eq. (67) possesses a trivial solution for $A\left( r\right) $ irrespective
of the form of $F\left( r\right) $ which is given by%
\begin{equation}
A\left( r\right) =C_{0}r^{2}
\end{equation}%
in which $C_{0}$ is an integration constant. In this situation the field
equations are all satisfied provided $B\left( r\right) $ and $A\left(
r\right) $ satisfy the condition (68) i.e.,%
\begin{equation}
B=\frac{C_{0}r^{2}e^{4\int \frac{F^{\prime }}{rF^{\prime }+F}dr}}{\left(
rF^{\prime }+F\right) ^{2}}.
\end{equation}%
One can see easily that with $F=\xi =$constant we obtain 
\begin{equation}
B=\frac{C_{0}}{\xi ^{2}}r^{2}.
\end{equation}%
The solution given by (69), (71) and 
\begin{equation}
f\left( R\right) =\eta +\xi R
\end{equation}%
with $\eta =const.,$ constitutes a particular class. Note that since the
choice of $F\left( r\right) $ in (70) is arbitrary this can be used to
generate an infinite class of solutions. This will not be searched any
further here.

\subsubsection{$3+1-$dimensional black hole solution in $f\left( R\right)
=R+2\protect\alpha \protect\sqrt{R}$ gravity}

Previously we found an exact solution for the model of gravity in the form 
\begin{equation}
f\left( R\right) =R+2\alpha \sqrt{R}
\end{equation}%
with $\alpha \neq 0.$ The solution to the field equations is given by%
\begin{equation}
A\left( r\right) =B(r)=\frac{1}{2}+\frac{1}{3\alpha r}
\end{equation}%
with 
\begin{equation}
R=\frac{1}{r^{2}}.
\end{equation}%
The solution is a black hole solution with $\alpha <0$ and therefore one may
write%
\begin{equation}
ds^{2}=-\frac{1}{2}\left( 1-\frac{r_{+}}{r}\right) dt^{2}+\frac{dr^{2}}{%
\frac{1}{2}\left( 1-\frac{r_{+}}{r}\right) }+r^{2}d\Omega ^{2}
\end{equation}%
in which the horizon is shown as $r_{+}.$ A change of variables of the form $%
t=\sqrt{2}T,$ $r=\frac{\rho }{\sqrt{2}}$ reduces (76) to%
\begin{equation}
ds^{2}=-\left( 1-\frac{\rho _{+}}{\rho }\right) dT^{2}+\frac{d\rho ^{2}}{1-%
\frac{\rho _{+}}{\rho }}+\frac{1}{2}\rho ^{2}d\Omega ^{2}
\end{equation}%
which is the Schwarzschild black hole with a deficit angle caused by a
cosmic string.

\section{Generalization to $f\left( R\right) $ gravity coupled to matter
sources}

In this Section we extend our vacuum analysis to the presence of matter
coupled with gravity. Therefore the action becomes%
\begin{equation}
I=\int d^{D}x\sqrt{-g}\left( \frac{1}{16\pi G}f\left( R\right) +\mathcal{L}%
_{m}\right)
\end{equation}%
in which $\mathcal{L}_{m}$ is the matter Lagrangian density. The field
equations become%
\begin{equation}
FR_{\mu }^{\nu }-\frac{1}{2}f\delta _{\mu }^{\nu }-\nabla ^{\nu }\nabla
_{\mu }F+\delta _{\mu }^{\nu }\square F=8\pi GT_{\mu }^{\nu }
\end{equation}%
with $T_{\mu }^{\nu }=diag\left( -\rho ,p,q,q\right) $ the energy momentum
tensor of the matter source. The line element is going to be a spherically
symmetric as (1) and without going through the details of the field
equations, we give the changes in the field equations (11) and (13). The
corresponding field equation to Eq. (11) reads%
\begin{equation}
A=BH^{2}\exp \left( -2\int \left( \frac{D_{1}F^{\prime }}{H}-\frac{\left(
\rho +p\right) }{BH}\right) dr\right)
\end{equation}%
while the main equation (13) i.e. the master equation for $B\left( r\right) $
takes the form%
\begin{equation}
B^{\prime \prime }+\left( P+\frac{r\left( p+\rho \right) }{HB}\right)
B^{\prime }+2QB+2S+\frac{2r^{2}\left( \rho +p\right) }{H^{2}B}=0
\end{equation}%
in which $P$, $Q$ and $H$ are given in (14), (15) and (12), respectively and%
\begin{equation}
S=\frac{D_{3}}{r^{2}}+\frac{\left( p-q\right) }{F}-\frac{\left(
D_{1}F^{\prime }-rF^{\prime \prime }\right) \left( \rho +p\right) }{H^{2}}.
\end{equation}%
Note that, unlike the source-free case, here the master $B$ equation is not
a linear equation. It can also be observed that the particular choice of $%
\rho +p=0$ removes the non-linearity in the $B$ equation (81). Further
choice of $p=q$ leaves us with the same equation for $B$ as in the
sourceless case. Yet with $q\neq 0,$ the source shows itself in the $\Psi $
function as given in the sequel.

The closed form of $f$ is given by%
\begin{equation}
f=-2q+\frac{1}{r^{2}A}\Psi
\end{equation}%
where%
\begin{multline}
\Psi =2r^{2}ABF^{\prime \prime }+\left[ BrA^{\prime }+\left( rB^{\prime
}+2BD_{3}\right) A\right] rF^{\prime } \\
-F\left[ BrA^{\prime }+\left( rB^{\prime }+2\left( B-1\right) D_{3}\right) A%
\right] .
\end{multline}%
The foregoing expressions are not impressive much unless we provide concrete
examples. This is our aim in the section that follows.

\subsection{Applications}

\subsubsection{$3+1-$dimensional global monopole coupled to $f=R+2\protect%
\alpha \protect\sqrt{R}$ modified gravity}

One immediate application of Eq. (81) is the extension of the $f=R+2\alpha 
\sqrt{R}$ in $3+1-$dimensional vacuum to the gravity coupled to the global
monopole \cite{54} whose energy momentum at very large distance is given by
(we assume $A=B$ in the line element)%
\begin{equation}
T_{\mu }^{\nu }=diag.\frac{\eta ^{2}}{r^{2}}\left[ 1,1,0,0\right]
\end{equation}%
in which $\eta $ represents the global monopole charge. The solution for the
metric is given by%
\begin{equation}
A=B=\frac{1}{2}+\frac{\left( 2\eta ^{2}+1\right) }{3\alpha r}
\end{equation}%
with 
\begin{equation}
R=\frac{1}{r^{2}}
\end{equation}%
and%
\begin{equation}
F\left( r\right) =1+\alpha r.
\end{equation}%
Again we must have $\alpha \neq 0$ and in the limit $\eta =0$ one recovers
the vacuum solution.

\subsubsection{$2+1-$dimensional Maxwell electric field coupled to $f(R)$
gravity}

Let's consider now Maxwell electric field to be coupled to $f\left( R\right) 
$ gravity in $2+1-$dimensions. Using the line element (1) and the standard
Maxwell Lagrangian together with $F=1+\alpha r$ one finds%
\begin{equation}
T_{\mu }^{\nu }=diag.\frac{Q^{2}}{r^{2}}\left( -1,-1,1\right) 
\end{equation}%
in which $Q$ is the electric charge. The solution with small $\alpha $ up to
first order simply reads%
\begin{multline}
A=B\simeq 2Q^{2}\left( 2\alpha r-1\right) \ln r \\
+2\alpha \left( Q^{2}+M\right) r-M+C_{2}r^{2}+\mathcal{O}\left( \alpha
^{2}\right) 
\end{multline}%
and%
\begin{equation}
f\simeq R+2C_{2}+\frac{4Q^{2}\alpha }{r}+\mathcal{O}\left( \alpha
^{2}\right) 
\end{equation}%
with%
\begin{multline}
R\simeq -6C_{2}+\frac{2Q^{2}}{r^{2}} \\
-\frac{4\alpha }{r}\left( 4Q^{2}+M+2Q^{2}\ln r\right) +\mathcal{O}\left(
\alpha ^{2}\right) .
\end{multline}%
Clearly by setting $\alpha =0$ one recovers the charged BTZ black hole with
cosmological constant $\Lambda =C_{2}.$ We add that the solution when the
energy momentum tensor of the matter source is of the form of a fluid, one
should consider it as an interior solution. Therefore one has to make sure
that the Israel junction conditions are satisfied at the interface between
exterior and interior solutions \cite{57}. In two examples we studied here,
the energy momentum tensors are long range fields which allow us to consider
our solutions to be exterior.

\section{Conclusion}

The integrability of $f\left( R\right) $ vacuum gravity with spherical /
circular symmetry is reduced first to a set of master relations, i.e., Eq.
(11) and a master equation for $B\left( r\right) ,$ i.e., Eq. (13). Given
any ansatz generating function $F=\frac{df}{dR}$ in terms of the coordinate $%
r,$ our method generates a solution pair of $A\left( r\right) ,$ $B\left(
r\right) $ and a solution for the function $f\left( R\right) $. Most of our
solutions encountered are of hybrid nature, that is $f\left( R\right) $
can't be expressed explicitly in terms of $R.$ The power-law form for
instance, of the form $f\left( R\right) \sim R^{k}$, with a rational number $%
k$, is obtained easily in our method. Some of the solutions presented as
applications are already known. Yet, new and rare type of solutions can also
be obtained easily. In the second part of the paper, we extend the
integrability of vacuum case to the non-vacuum $f\left( R\right) $ theories.
For this case we found also a master relation, i.e., Eq. (80) and a master
equation, i.e., Eq. (81). Considering $F\left( R\right) $ to be the
generating function for our formalism and $T_{\nu }^{\mu }=diag\left[ -\rho
,p,q,q\right] $ to be our energy momentum tensor one finds a solution to the
master equation (81). We presented two examples. In the first example we set 
$F=1+\alpha r$ with a global monopole coupled to the gravity in $3+1-$%
dimensions. The second example considers the Maxwell electric field coupled
to the gravity in $2+1-$dimensions with the same generating function. In
this case we found the solutions approximately for small $\alpha $. Let's
also add that among the few examples we have studied we found some closed
form of $f\left( R\right) $ such as $f\left( R\right) =R+2\alpha \sqrt{%
R-2\Lambda }-2\Lambda $ and $f\left( R\right) =\bar{\alpha}R^{1-\frac{a}{2}%
\text{ }}.$ We found that the first one can not satisfy the conditions for
the absence of ghosts and thermodynamic stability simultaneously while the
second one for specific $a$ does satisfy. Finally it will be in order to
state that our method of reduction for the spherically symmetric $f\left(
R\right) $ gravity has a large scope as far as solutions are concerned.
Physical implications of the solutions obtained such as dark matter
connection is not considered in the present study. Extension of our
formalism to problems with different symmetries, such as stationary axial
symmetry requires a separate investigation.

\bigskip

\bigskip

\end{document}